\documentclass{mpe_report}

\usepackage{psfig,graphicx,epsfig}
\usepackage{color}
\usepackage{amsmath,amssymb,epic,eepic,array}
\usepackage{natbib,aas_macros} 

\begin{document}

\pagenumbering{arabic}
\setcounter{page}{201}

 \renewcommand{\FirstPageOfPaper }{201}\renewcommand{\LastPageOfPaper }{204}
\title{Long term spectral variability in the soft gamma-ray repeater SGR\,1900+14}
\author{Paolo Esposito\inst{1,2} \and Sandro Mereghetti\inst{2} \and Andrea Tiengo\inst{2} \and Diego G\"otz\inst{3} \and Lara Sidoli\inst{2} \and Marco Feroci\inst{4}}
\institute{Universit\`a di Pavia, Dipartimento di Fisica Nucleare e Teorica and INFN-Pavia, via Bassi 6, I-27100 Pavia, Italy
\and  INAF-IASF Milano, via Bassini 15, I-20133 Milan, Italy
\and CEA Saclay, DSM/DAPNIA/Service d'Astrophysique, F-91191,
Gif-sur-Yvette, France
\and INAF-IASF Roma, via Fosso del Cavaliere 100, I-00133 Roma,
Italy }
\maketitle

\begin{abstract}
We present a systematic analysis of all the {\it BeppoSAX} data of SGR\,1900+14. 
The observations spanning five years show that the source was brighter than usual on two occasions: 
$\sim$20 days after the August 1998 giant flare and during the 10$^5$\,s long X--ray afterglow 
following the April 2001 intermediate flare. In the latter case, we explore the possibility of 
describing the observed short term softening only with a change of the temperature of a 
blackbody-like component. In the only {\it BeppoSAX} observation performed before the giant 
flare, the spectrum of the SGR\,1900+14 persistent emission was significantly harder and possibly 
detected also above 10 keV with the PDS instrument. In the last {\it BeppoSAX} observation 
(April 2002) the flux was $\sim$25\% lower than the historical level, suggesting that the source 
was entering a quiescent period. 
\end{abstract}

\section{Introduction}
\label{intro} The soft gamma-ray repeater (SGR) SGR\,1900+14 was discovered in 1979 through series of short and soft gamma-ray bursts \citep{mazets79}. Many years later, its persistent pulsating X--ray counterpart was discovered in the 2--10 keV energy band \citep{hurley99}. More recently, it was also detected in the hard X--ray range (\mbox{20--100 keV}) with the \emph{INTEGRAL} satellite, becoming the second SGR, after SGR\,1806$-$20, established as a persistent hard X-ray source \citep{gotz06}.\\
\indent The rather discontinuous bursting activity of SGR\,1900+14 (see \mbox{Fig.~\ref{fig:1}}, bottom panel) raised to a summit on 1998 August 27 with the emission of a giant flare, when more than 10$^{44}$ ergs of $\gamma$--rays were emitted in less than one second \citep{hurley99gf}. This was one of the three giant flares detected up to now from three different SGRs. The extreme properties of these events are the main motivation for the {\it magnetar} interpretation. In this model \citep{thompson95,thompson96}, the SGRs and the Anomalous X--ray Pulsars (AXPs, another class of X--ray sources with similar properties, see e.g. \citealt{mereghetti02}) are believed to be neutron stars powered by the decay of their extremely intense magnetic field (\mbox{$B$$\sim$10$^{14}$--10$^{15}$ G}) rather than by rotation.\\
\indent Here we present the analysis of the persistent emission of SGR\,1900+14 by means of the {\it BeppoSAX} satellite, both in the soft and hard X--ray range, and its evolution across the giant flare and in relation to its bursting activity.
\section{Soft X--ray emission}\label{sec:1}
\subsection{Observations and data analysis}\label{sec1.1} 
We have analyzed all the X--ray observations of SGR\,1900+14 performed with the {\it BeppoSAX} satellite (see Table \ref{tab:1}).
\begin{table}
      \caption{Log of the {\it BeppoSAX} observations of SGR~1900+14}
         \label{tab:1}
      \[
         \begin{array}{p{0.1\linewidth}cccc}
            \hline
            \noalign{\smallskip}
            Obs & \textrm{Date} & \multicolumn{3}{c}{\textrm{Instrument\,/\,exposure}}\\
            \noalign{\smallskip}
            \hline
            \noalign{\smallskip}
A& 1997\,\textrm{May}\,12 & \textrm{LECS}\,/\,20 \,\textrm{ks} & \textrm{MECS}\,/\,46 \,\textrm{ks} & \textrm{PDS}\,/\,20\,\textrm{ks} \\
B& 1998\,\textrm{Sep}\,15 & \textrm{LECS}\,/\,14 \,\textrm{ks}  & \textrm{MECS}\,/\,33 \,\textrm{ks} & \textrm{PDS}\,/\,16 \,\textrm{ks} \\
C& 2000\,\textrm{Mar}\,30 & \textrm{LECS}\,/\,14 \,\textrm{ks}  & \textrm{MECS}\,/\,40 \,\textrm{ks} & \textrm{PDS}\,/\,18 \,\textrm{ks} \\
D& 2000\,\textrm{Apr}\,25 & \textrm{LECS}\,/\,17 \,\textrm{ks}  & \textrm{MECS}\,/\,40 \,\textrm{ks} & \textrm{PDS}\,/\,19 \,\textrm{ks} \\
E& 2001\,\textrm{Apr}\,18 & \textrm{LECS}\,/\,20 \,\textrm{ks}  & \textrm{MECS}\,/\,46 \,\textrm{ks} & \textrm{PDS}\,/\,17 \,\textrm{ks} \\
F& 2001\,\textrm{Apr}\,29 & \textrm{LECS}\,/\,26 \,\textrm{ks}  & \textrm{MECS}\,/\,58 \,\textrm{ks} & \textrm{PDS}\,/\,26 \,\textrm{ks} \\
G& 2002\,\textrm{Mar}\,09 &  $-$              & $-$             & \textrm{PDS}\,/\,48 \,\textrm{ks} \\
H& 2002\,\textrm{Apr}\,27 &  $-$              & \textrm{MECS}\,/\,83 \,\textrm{ks} & $-$            \\
            \noalign{\smallskip}
             \hline
         \end{array}
      \]
   \end{table} 
The spectra were extracted from the MECS \citep{boella97mecs} and LECS \citep{parmar97} instruments using circles with radii 4$'$ and 8$'$, respectively. The background spectra were extracted in all cases from nearby regions and time filters were applied to both the source and background spectra to exclude the SGR bursts detected during observation B, E and F. 
\subsection{Spectral results}\label{specres}

\begin{figure}
\centering
  \includegraphics[width=11.0cm,angle=90]{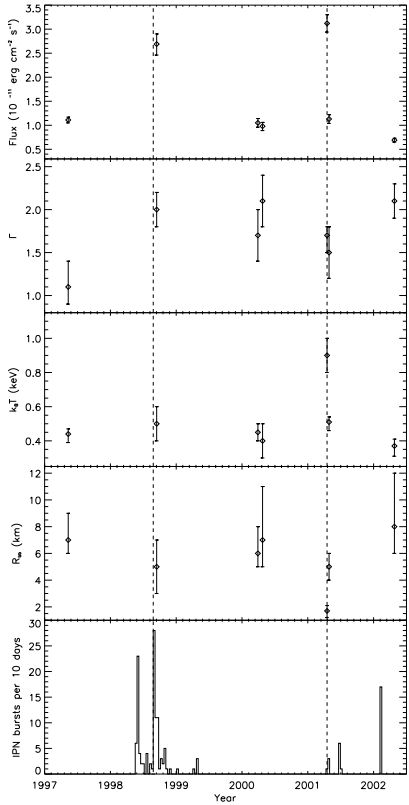}
\caption{Long term evolution of the 2--10 keV unabsorbed flux, the spectral parameters (for an absorbed power-law plus blackbody model, assuming $n_{\rm{\scriptscriptstyle{H}}}$=2.6$\times$10$^{22}$\,cm$^{-2}$) and the burst activity (as observed by the Interplanetary Network) of SGR\,1900+14. The vertical dashed lines indicate the giant and intermediate flares (1998 August 27 and 2001 April 18, respectively).} \label{fig:1}
\end{figure}

\begin{figure}
\centering
  \includegraphics[height=7.0cm,angle=90]{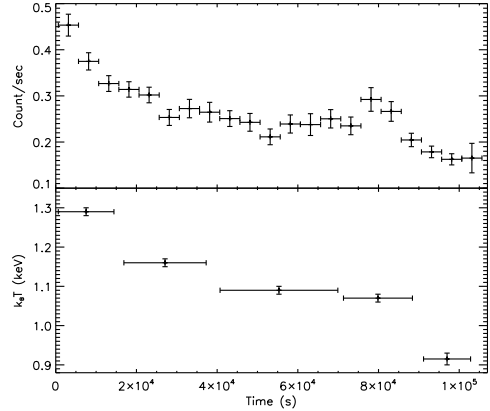}
\caption{Background subtracted MECS 2--10 keV light curve and blackbody temperature observed on 2001 April 18 about 7.5 hours after the flare. The latter values correspond to an additional blackbody component that can be interpreted as due to a portion of the neutron star surface heated during the flare (see Section \ref{specres}).} \label{ultimo}
\end{figure}

We have first tried to fit the spectra with an absorbed power-law model, but three observations give unacceptable values of the $\chi^2$ and structured residuals. For these observations, good fits are obtained with the addition of a blackbody component. Since such a two-components model is typical of the magnetar candidates \citep{woods04}, we have used this model to fit all the available spectra, obtaining the results reported in \mbox{Fig.~\ref{fig:1}}. The blackbody parameters are compatible in all the available observations, except for that taken during the afterglow of the intermediate flare (observation E). This indicates that a constant blackbody component with \mbox{$k_BT$$\sim$0.4 keV} and emitting area with \mbox{$R$$\sim$6--7 km}, might be a permanent feature of the X-ray spectrum of SGR\,1900+14.\\
\indent As can be seen in  the upper panel of \mbox{Fig.~\ref{fig:1}}, the flux varies by a factor $>$5, with the highest values observed during observations B and E. These two observations were taken shortly after extreme bursting events. The former was performed 20 days after the giant flare, that was followed by a $\sim$2 months period of enhanced X--ray flux \citep{woods99b}. The latter started only 7.5 hours after the intermediate flare which had a fluence $\sim$20 times lower than that of the 2001 April 18 giant flare  \citep{guidorzi04}.\\
\indent The afterglow following this bright burst is clearly visible during the {\it BeppoSAX} observation as a decrease in the X--ray flux (see \mbox{Fig.~\ref{ultimo}}, top panel), accompanied by a significant softening of the spectrum  \citep{feroci03}. In addition to the afterglow analysis already reported by \citet{feroci03}, we have performed a time resolved spectroscopy of the afterglow by dividing observation E into five time intervals. Under the assumption that the variable ``afterglow'' emission is present on top of a ``quiescent'' emission that shows only moderate variations on long time-scales, we fitted them with a model consisting of a power-law plus blackbody with fixed parameters (as representative parameters of the fixed quiescent emission we used values consistent with those seen in the last observations before the flare: C and D), plus a third variable component to model the afterglow emission. Although the ``afterglow emission'' in the five spectra can be fitted by a variety of models, the spectral evolution of the afterglow is well represented by an additional blackbody component with fixed emitting area (radius of $\sim$1.5 km, for a source distance of 15 kpc) and progressively decreasing temperature ($k_BT$ from $\sim$1.3 to  $\sim$0.9 keV, see \mbox{Fig.~\ref{ultimo}}, bottom panel), that can be interpreted as due to a portion of the neutron star surface heated during the flare.
\\
\indent Excluding the two observations taken after the exceptional explosive events (B and E), the flux of SGR\,1900+14 had a rather constant value of $\sim$$10^{-11}$ erg cm$^{-2}$ s$^{-1}$ from 1997 to 2001. On the other hand a significantly lower flux level was seen in the following observations. The flux decrease actually started when the source was still moderately active (the flux in observation H is at least $\sim$20\% lower than in all the previous quiescent observations) and has been interrupted by a slight rise in coincidence with the March 2006 burst reactivation, as shown by recent {\it XMM-Newton} observations \citep{met06}.\\
\indent Although the flux of the only pre-giant flare observation is compatible with that of the quiescent post-flare observations taken before 2002, its spectrum is significantly harder, as shown in the second panel of \mbox{Fig.~\ref{fig:1}}. The overall hardness of the pre-flare observation is confirmed by the fact that the spectra C, D, F, and H can be simultaneously fit with the same parameters (introducing a normalization factor to account for the flux change), while the addition of spectrum A gives an unacceptable fit with structured residuals.
\section{Hard X--ray emission} \label{sec:2}
\subsection{Detection with the PDS instrument}
To study the high energy emission from SGR\,1900+14 we used the {\it BeppoSAX} PDS instrument \citep{frontera97}, which operated in the 15--300 keV range. The PDS instrument was more sensitive than \emph{INTEGRAL} in this energy band, but it had no imaging capabilities and therefore the possible contamination from nearby sources must be taken into account. The field of view of the PDS instrument was 1.3$^{\circ}$ (FWHM) and the background subtraction was performed through a rocking system that pointed to two 3.5$^{\circ}$ offset positions every 96 s.\\
\indent In the case of SGR\,1900+14, the background pointings were free of contaminating sources, as confirmed by the identical count rates observed in the two offset positions during each observation. The field of SGR\,1900+14 is instead rather crowded, with three transient sources: the X--ray pulsars 4U~1907+97 \citep{liu00} and XTE~J1906+09 \citep{liu00}, and the black hole candidate XTE~J1908+94 \citep{intzand02}, located at angular distances of 47$'$, 33$'$ and 24$'$ from the SGR, respectively. The pulsations of the two pulsars are clearly visible in the PDS data below 50 keV when they are active, while XTE~J1908+94, if in outburst, is clearly visible in the simultaneous MECS and LECS images and, being very bright, also in the lightcurve collected by the All Sky Monitor (ASM) on board the {\it RossiXTE} satellite. We have found that at least one of these contaminating sources was on in all the {\it BeppoSAX} observations except for the first one (see Table \ref{tab:2}). 
Thus, only the 1997 observation (observation A), during which a significant signal was detected in the background subtracted PDS data, can be used to study SGR\,1900+14 without the problem of contaminating sources. We searched for the SGR pulsation period (5.15719 s, as measured in the simultaneous MECS data) in the PDS data, but the result was not conclusive, giving only a 3$\sigma$ upper limit of 50\% to the pulsed fraction of a sinusoidal periodicity, to be compared to the $\sim$20\% pulsed fraction observed below 10 keV.
\begin{table}
      \caption{Status of the three transient sources within the PDS field
of view during the {\it BeppoSAX} observations. The presence of the
two X--ray pulsars (4U~1907+97 and XTE~J1906+09) is confirmed by the presence of their pulsations in the PDS data, while the black hole
candidate (XTE~J1908+94) by its detection in the MECS and LECS images and in
the {\it RossiXTE} ASM lightcurve.}
        \label{tab:2}
      \[
         \begin{array}{p{0.2\linewidth}ccc}
            \hline
            \noalign{\smallskip}
            Obs & \textrm{4U~1907+97} & \textrm{XTE~J1906+09} &  \textrm{XTE~J1908+94}\\
            \noalign{\smallskip}
            \hline 
	    \noalign{\smallskip}
         A & \textrm{OFF} & \textrm{OFF} & \textrm{OFF}\\
B & \textrm{ON} & \textrm{ON} & \textrm{OFF} \\
C & \textrm{OFF} & \textrm{ON} & \textrm{OFF} \\
D & \textrm{ON} & \textrm{ON} & \textrm{OFF} \\
E & \textrm{OFF} & \textrm{ON} & \textrm{OFF} \\
F & \textrm{ON} & \textrm{ON} & \textrm{OFF} \\
G & \textrm{OFF} & \textrm{ON} & \textrm{ON} \\
            \noalign{\smallskip}
             \hline
         \end{array}
      \]
   \end{table} \\
\indent Although we cannot rule out contamination from unknown transient sources, the flux measured in observation A up to $\sim$150 keV is most likely due to SGR\,1900+14 and therefore hereafter we assume that the spectral properties discussed in the text are associated to SGR\,1900+14 alone.
\subsection{Spectral analysis}
\begin{figure}
\centering

\includegraphics[width=6.5cm,angle=-90]{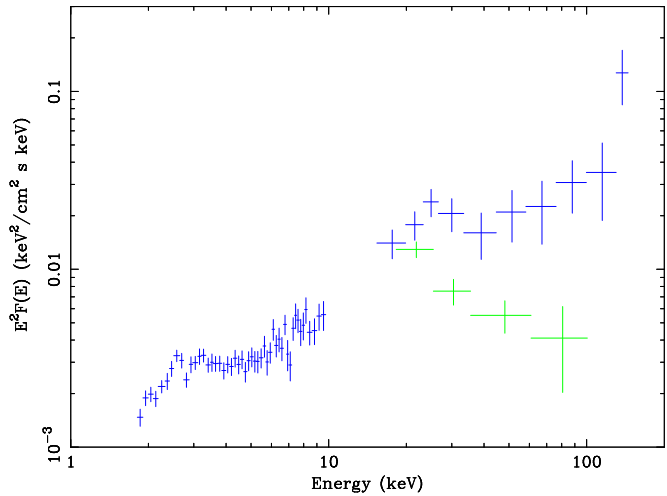}
\caption{Blue points: broad band spectrum of \mbox{SGR\,1900+14} taken on 1997 May 12 (observation A) with {\it BeppoSAX} (both MECS and PDS data). Green points: \emph{INTEGRAL} data from March 2003 to June 2004.} \label{fig:2}
\end{figure}

The background subtracted PDS spectrum of observation A can be well fit by a power-law with photon index $\Gamma$=1.6$\pm$0.3, significantly flatter than that measured by \emph{INTEGRAL} (\mbox{$\Gamma$=3.1$\pm$0.5}, see \mbox{Fig.~\ref{fig:2}}) in 2003\,/\,2004. The corresponding 20--100 keV flux is \mbox{6$\times$10$^{-11}$ erg cm$^{-2}$ s$^{-1}$}, a factor $\sim$4 higher than during the \emph{INTEGRAL} observations, which confirms that before the giant flare the hard X--ray tail of \mbox{SGR\,1900+14} was brighter.\\
\indent The \emph{INTEGRAL} spectrum was collected during \mbox{$\sim$2.5 Ms} of different observations performed between March 2003 and June 2004, and thus it represents the hard X--ray emission of \mbox{SGR\,1900+14} averaged over that long time period. Therefore, its relation to the soft X--ray spectrum can be studied only comparing the spectra taken by other instruments in a similar time period, as shown for example in \mbox{Fig.~\ref{fig:2}}. The PDS instrument, instead, being a high sensitivity hard X--ray detector coupled to the MECS and LECS soft X--ray cameras, gives us the chance to study the broad band spectrum of \mbox{SGR\,1900+14} during a single observation. Fitting the \mbox{1--150 keV} {\it BeppoSAX} spectrum of observation A, we obtain a good result ($\chi^2$=1.17 for 136 degrees of freedom) simply extrapolating to higher energies the best-fit model found in the soft X--ray range. In fact, a fit with an absorbed power-law plus blackbody model gives the following parameters: photon index $\Gamma$=1.04$\pm$0.08, blackbody temperature $k_BT$=0.50$\pm$0.06, radius $R_{bb}$=5$\pm$2 km, and absorption $n_{\rm{\scriptscriptstyle{H}}}$=(1.8$\pm$0.5)$\times$10$^{22}$ cm$^{-2}$.

\section{Conclusions} \label{sec:3}
We have studied the variability of \mbox{SGR\,1900+14}, both in the hard and in the soft X--ray range, finding the following results:
\begin{itemize}
\item Except for the observations immediately following exceptional flares, the flux level in soft X--rays was stable while the source was moderately active and progressively decreased when it entered a 3 years long quiescent period.
\item The intermediate flare of 2001 April 18 was followed by an X--ray afterglow that can be interpreted as due to the heating of a significant fraction of the neutron star surface, that then cools down in $\sim$1 day. This is consistent with the interpretation of similar events in other magnetar candidates
\citep{woods04axp}.
\item The soft X--ray spectrum during the only available pre giant flare
observation was harder than in the following quiescent observation.
This is similar to what observed in SGR~1806--20, the only other SGR
that could be monitored before and after a giant flare \citep{mte05,tiengo05}.
\item Comparing the hard X-ray spectrum of \mbox{SGR\,1900+14} recently observed with \emph{INTEGRAL} to that observed with the PDS instrument in 1997, we find evidence for variations in flux and spectral slope.
\item The reduction of the X--ray tail in coincidence with the giant flare is supported by the count rates detected in the PDS instrument above 50 keV during the different {\it BeppoSAX} observations, that indicate how they significantly decreased already in the first post-flare observation. Since the hard X--ray tail in the spectrum of \mbox{SGR\,1900+14} might contain most of its total emitted energy, its variability in relation to the bursting activity is a key point to try to understand the SGR emission processes.
\end{itemize}
\begin{acknowledgements}
We gratefully acknowledge the support by the WE-Heraeus foundation.
\end{acknowledgements}
\bibliographystyle{aa}
\bibliography{biblio}


          \clearpage

\end{document}